\documentclass{article}
\usepackage{caption}
\usepackage{apacite}
\usepackage{amsmath}
\usepackage{amssymb}
\usepackage{graphicx}

\usepackage{epsfig}

\title{The Cultural Evolution of National Constitutions}

\author
{Daniel N. Rockmore,$^{1,2,3,4\ast}$ 
Chen Fang,$^{1}$
Nicholas J. Foti,$^{5}$\\
Tom Ginsburg,$^{6}$
David C. Krakauer$^{3}$\\
\\
\normalsize{$^{1}$Department of Computer Science, Dartmouth College, Hanover, NH, USA 03755}\\
\normalsize{$^{2}$Department of Mathematics, Dartmouth College, Hanover, NH, USA 03755}\\
\normalsize{$^{3}$The Santa Fe Institute, Santa Fe, NM, USA 87501}\\
\normalsize{$^{4}$The Neukom Institute for Computational Science, Dartmouth College, Hanover, NH, USA 03755}\\
\normalsize{$^{5}$Department of Statistics, University of Washington, Seattle, WA USA 98195-4322}\\
\normalsize{$^{6}$University of Chicago Law School, The University of Chicago, Chicago, IL, USA  60637}\\
\\
\normalsize{$^\ast$To whom correspondence should be addressed; E-mail:  rockmore@math.dartmouth.edu.}
}

\date{\today}

\begin{document} 


\baselineskip18pt

\maketitle


\begin{abstract}
 We explore how ideas from infectious disease and genetics can be used to uncover patterns of cultural inheritance and innovation in a corpus of 591 national constitutions spanning 1789--2008. Legal ``Ideas" are encoded as ``topics" - words statistically linked in documents -  derived from topic modeling the corpus of constitutions. Using these topics we derive a diffusion network for borrowing from ancestral constitutions back to the US Constitution of 1789 and reveal that constitutions are complex cultural recombinants. We find systematic variation in patterns of borrowing from ancestral texts and ``biological"-like behavior in patterns of inheritance with the distribution of ``offspring" arising through a bounded preferential-attachment process. This process leads to a small number of highly innovative (influential) constitutions some of which have yet to have been identified as so in the current literature. Our findings thus shed new light on the critical nodes of the constitution-making network. The constitutional network structure reflects periods of intense constitution creation, and systematic patterns of variation in constitutional life-span and temporal influence. 
\end{abstract}

\section*{Introduction}

Cultural inheritance involves the diffusion of innovations, a process  of interest to both biologists \cite{Hartl}  and social scientists \cite{ERogers}. In biology inheritance is governed by mechanisms of genetic transmission, which have been quantified \cite{Christiansen}.  Cultural inheritance takes a variety of forms which can resemble variants of biological inheritance \cite{sforza, Richerson, Mesoudi}, including cultural selection \cite{Rogers, Pagel}. In cultural domains, complex forms of knowledge are encoded in social norms, legal principles and scientific theories \cite{Wimsatt, Kaiser} and follow complex forms of transmission that involve the coordinated borrowing and learning of constellations of ideas, producing a diversity of phylogenetic patterns \cite{mace}.

Now that a large body of the cultural record has been digitized (including books \cite{googlebooks}, music \cite{imslp}, art \cite{artstor}, etc.)  new techniques of machine learning  are making the quantitative analysis of high-dimensional cultural artifacts possible.  In analogy with the biological sciences, and genetics in particular, this data mining approach to the analysis of culture is sometimes referred to as ``culturomics" \cite{pinkerscience}, a term born of  the consideration of  the frequency distribution of an $n$-gram  in the Google Books corpus over time \cite{googlebooksngramviewer}  as proxy for how memes move in and out of the cultural record. Literature (and text generally) remains a primary focus of such work  (see e.g., \cite{Moretti2005,jockers,pnas-lit}). A fascinating challenge is to supplement these correlation-based approaches to the understanding of cultural evolution with principled causal mechanisms directed at discovering fundamental, extra-biological evolutionary processes.

We consider the notion of {\em diffusion patterns} in the study of cultural inheritance as a means of tracking the diffusion of {\em topics} through the documents in a legal text corpus of five hundred and ninety-one national constitutions (the  full list is given in the Supplementary Materials Table S1).   
``Topics" has a technical meaning here (and throughout this paper that is the sense in which the word is used) as probability distributions over words (positive weights that sum to one) that are the output of  {\em topic modeling}, which is a computational and statistical methodology for  text analysis that has made great inroads throughout the humanities (see e.g., \cite{riddell}),  to the point of reaching an almost ``plug-and-play" form (see e.g., \cite{topicmodeling1}) for easy deployment.  A set of topics is ``learned" (i.e., automatically derived) from the corpus. The various topic distributions  highlight (i.e., attach high weight to) different sets of words. In the best cases those words usually suggest a particular theme and associated labeling of the topic. Texts in the corpus are partitioned into chunks, which are  thus represented as varying weighted mixtures of topics. In this way  topics provide a low-dimensional representation of the corpus in terms of higher level ideas and  provide a rigorous operational basis for a meme, to be tested against a suitable dynamics of inheritance. Although we focus on its use in the analysis of text, the topic modeling framework  is more general and has been used in a number of areas \cite{Blei-ACM}.

Given a topic of some significance in a work, embodied in a set of semantically correlated legal concepts, we track its appearance and prevalence in subsequent constitutions within the corpus, as well as its extinction. While dynamical considerations  have been incorporated previously into topic models \cite{Blei:2006:DTM, Wang:2006:TOT} this analysis differs in that we account for the diffusion of topics from document to document, and in this way reveal more clearly the patterns of genealogy and the essentially recombinant nature of textual artifacts. These resemble in the parallel domain of invention the recombinant quality of patents \cite{Youn}. It is our contention that while culture is clearly an active \textit{in situ} feature of human brains \cite{Boyd}, it is also present in material artifacts which afford rich forms of combinatorial manipulation and transmission \textit{ex situ}.

The corpus of national constitutions is particularly well-suited to a framing and analysis as a document corpus composed of units of correlated meaning evolving according to idea diffusion and borrowing.  Indeed, scholars have demonstrated that many provisions in constitutions are copied from those of other countries. For example, through $n$-gram analysis Ginsburg et al.  \cite{FGR-preambles} show that constitutional preambles, which are conceptualized as the most nationally localized part of constitutions, also speak in a universal idiom and include a good deal of borrowing. Law and Versteeg \cite{LawVersteeg7} have shown that rights provisions have spread around the globe.  Elkins et al. \cite{ElGinsMel8,ElGinSim8-2013} show that some rights, such as freedom of expression, have become nearly universal, while others have not.  Some even argue that there is a kind of global script at work, whereby nation-states seek to use constitutions to participate in global discourses  \cite{Go9,BoliBennett9,Law9}.
This evolutionary framing of the creation of national constitutions draws on broader biological analogies for legal development across time and space \cite{Watson11}. Our use  of {\em diffusion trees} as a framework for the study of this problem (see the Methods section in the Supplementary Materials for details) can be seen as a novel quantification of this biological analogy.

It is important that we are clear that this integration of topic modeling and diffusion networks enables only a quantitative  articulation and tracking of instances of thematic similarity over time. The links we demonstrate across texts are consistent with a model in which one text influences another.  However, our approach does not demonstrate the specific mechanisms by which influences are transmitted, so we focus instead on the sequential patterns in which textual material flows across time and space. As we demonstrate in our Discussion, this enables an analysis enhancing traditional scholarly opinion as regards the usual notion of ``influence",  while also at times uncovering   temporal connections suggesting further or new investigations.

\section*{Results}

As mentioned, a {\em topic} is a probability distribution over a fixed {\em vocabulary} derived from a text {\em corpus}.  It thus represents a correlated set of words  encoding something like a ``meme" or stochastic set of associations. (Technically, the pre-processing of the texts may result in some elements of the vocabulary set that are not words per se, but instead word stems, often called ``tokens". We will use the more colloquial term ``word" in this paper.)  The text corpus is partitioned into {\em documents}, sets of roughly contiguous groupings of $500$ words. This is a standard topic modeling document length, short enough to reflect local context and long enough to make sensible the statistical model. In the best case each constitution would be partitioned into contiguous word-blocks, but processing may remove the odd abbreviation, title, etc. besides respecting natural boundaries, such as the end of one constitution and the beginning of another.   In the case of our corpus of constitutions, each constitution generally comprises a subset of such documents. The model does not take into account word order, just which words occur and in what frequencies. This is the so-called ``bag-of-words" model or representation, which is then encoded as a probability distribution over the vocabulary (the frequencies are positive and sum to one). 

{\em Topic modeling} is a methodology for learning topics such that each document (represented as a bag of words) is represented as a weighted sum (mixture) of topics. In its generative form, the topic model encodes the creation of each document by first choosing a topic according to the mixture of topics that the document comprises and then choosing a word according to the distribution of that particular topic. In this respect a constitution can be thought of as a ``meme cloud" with the topics encoding the memes.  We use the latent Dirichlet allocation (LDA) topic model (see ~\cite{BleiNJ03} for a discussion of  the various parameters that define the model). LDA is effectively the topic modeling industry standard. We tested several choices for the number of topics and chose $100$ which we then validated (cf. the Methods section in the Supplementary Materials for details).

The output of the topic model forms the basis for our results. They include (1) the discovery of the topics that make up the corpus of constitutions, (2) the determination of their flow through time (``information cascades"),  (3) the reconstruction of cultural diffusion trees;  (4) network analysis of diffusion trees; and (5) discovery of a very biological pattern of inheritance with a highly skewed pattern of cultural fertility. 

\subsection*{Topics} 
The $100$ topics were ``hand-labeled" by a constitution expert.  Note that ``hand-labeling" of topics is standard. Further elaboration on this can be found in our Discussion. Since generally each constitution comprises a set of corpus documents we assign an overall constitutional weight for a topic as the average topic weight over the documents that it comprises. In Table 1 we list the ten topics with largest average topic weight (over all the constitutions), along with the ten most probable (heavily weighted) words (in decreasing order) for each topic.\footnote{A full list of the topics, in order of average weight, with the weights of the top 20 words can be found at
 https://www.math.dartmouth.edu/$\sim$rockmore/topics\_weight\_order.txt.}

\begin{table*}[ht]
\begin{tabular*}{\hsize}{@{\extracolsep{\fill}}lc}
\textbf{Topic name}&\textbf{Top 10 words in topic}\cr
\hline
\hline
General rights &
right   
	rights  	
	citizens   	
	freedom   	
	law \\ &  	
	public   	
	guaranteed   		
	citizen everyone religious \\
\hline
Sovereignty&	national    	
	people    		
	sovereignty    		
	law    	
	rights\\ &    	
	state    		
	flag    	
	language international equal \\
\hline
public order&law    	
	public   		
	cases    		
	order    		
	one    \\ &		
	property     	
	laws    	
	authority    		
	liberty    		
	civil \\
\hline
separation of powers& congress    	
	executive    		
	laws    	
	power    		
	ministers  \\ &  	
	state    		
	secretaries    		
	order    		
	necessary    		
	public\\
\hline
organic law&law    	
	government    		
	president    	
	organization    		
	national    	\\ &	
	organic    	
	public    		
	laws    	
	social    		
	functioning  \\
\hline
socialism&people    	
	socialist    		
	country    		
	revolution   	
	working    \\ &	
	popular    		
	citizens   		
	system    		
	society    	
	development\\
\hline
legislative sessions&	session    		
	deputies    	
	sessions    	
	deputy    		
	members    \\ &		
	elected   		
	first    	
	vote    	
	majority    		
	extraordinary \\
\hline
bureaucracy&papers   		
	years    		
	state    		
	department    		
	necessary    	\\ &	
	respective    		
	individuals    		
	departments   	
	body    		
	power    \\
\hline
socialism legislature&people    		
	organs   		
	state    		
	supreme    		
	work    \\ &		
	organ    		
	presidium    	
	elected    		
	decisions    		
	committees \\
\hline

\end{tabular*}
\caption{Most popular topics across entire corpus, and their corresponding top 10 words. A full list of the topics, in order of average weight, with the weights of the top 20 words can be found at https://www.math.dartmouth.edu/$\sim$rockmore/topics\_weight\_order.txt.}
\end{table*}

\subsection*{Influence and clustering}

The identification of the topics now gives a natural way to represent a constitution as a mixture of probability distributions. With that, we can  compare quantitatively constitutions and  get at a quantitative notion of influence, completely driven by the data of the words. A first coarse pass at this is to create a constitutional ``family tree", where the (unique) immediate ancestor of any given constitution is simply the constitution closest to it among all earlier constitutions. Given that our constitutions are now represented as probability distributions (over topics), a natural measure of distance is the Kullback-Liebler (KL) divergence. Recall that the KL divergence of  probability distributions $P$ and $Q$ is defined as  $KL(P||Q) = \sum_i P(i)\log\frac{P(i)}{Q(i)}$.  KL is inherently non-symmetric. A standard interpretation\footnote{See e.g., https://en.wikipedia.org/wiki/Kullback-Leibler\_divergence.} is the degree to which a distribution $Q$  approximates another distribution $P$. So thinking of an earlier constitution as a potential model for a newly written constitution, the KL divergence of their underlying topic probability distributions is a natural measure of similarity.  

The ``KL Constitution Tree" is shown in Figure 1. Note that the figure is not scaled horizontally for time. The size and form of the representation presents some difficulty for reproducing legibly herein, so a separate pdf document, readily magnifiable, can be found online. {\footnote{See
https://www.math.dartmouth.edu/$\sim$rockmore/kl-tree.pdf.} We also include a detail.

The KL-tree is a coarse and aggregate articulation of the notion that constitutional ideas flow in time. It is also purely correlative and local. We should also like to explore global patterns of influence and the possibility of causal influence. We approach this by considering the ``flow'' of topics through constitutions and through time. Each instance of a topic flowing  appearing in a constitution (above some fixed threshold) is treated as a ``cascade''. We follow standard conventions \cite{CascadeRef} and define an {\em information cascade} as a collection of constitutions and their timestamps where each  topic in the constitution makes up a proportion greater than a robust threshold value.  When two constitutions (nodes) both express a topic above threshold then we consider this pair as a candidate for information ``cascading" from the earlier to the later. 

The topic cascades form the underlying data for a mode of  inference for  how ideas represented by topics are likely to have propagated through the corpus over time.  As stated previously, we view the observation of a topic (above some threshold) in two constitutions as a quantitative measure indicating correlation across time. Given the content of the topics and the fact that the constitutions are ordered chronologically and typically clustered spatially (see the Network Analysis subsection below and Figure 2), shared topics may very well have spread from the earlier to the latter, and hence are at least consistent with weak causality.  In order to learn the most likely propagation structure of the topics (given the data) we estimate an underlying {\em diffusion network} for the corpus~\cite{tkdd/Gomez-RodriguezLK12}.
A \textit{diffusion network} is a directed graph with nodes corresponding to constitutions and where the edges satisfy the condition that the source constitution predates the destination constitution. This imposes weak causal structure on the correlations.  Importantly, we do not observe the diffusion network, but only the cascades that are assumed to diffuse over it and are consistent with it.  In brief,  a probabilistic model describing the consistency of the observed cascades  with respect to a fixed diffusion network is defined.  The diffusion network is that which (approximately) maximizes this probability~\cite{tkdd/Gomez-RodriguezLK12}.

The presentation of the full diffusion tree on our corpus presents some visualization challenges. To give a sense of what it looks like, 
Figure S1
 shows  the entire learned diffusion tree on a restricted set of ninety-nine constitutions. Even this is too dense to be inspected visually for information, but the figure at least gives a good sense of the way in which the methodology reifies the phenomena of the idea diffusion. Each of the edges (directed and extending downward) indicate particular topics diffusing forward in time to be taken up by subsequent constitutions. Issues of readability make it impossible to put labels on the various edges. The optimization algorithm that produces the diffusion network only collects a subset of the topics that appear in a constitution. Some diffuse forward, others do not. The ``offspring" of a given constitution thus borrow certain ``ideas" of the parents, but others are created afresh, presumably depending on legally appropriate contextual factors. 

\subsection*{Network analysis}
In order to discern patterns in the diffusion tree the diffusion network is subjected to a clustering analysis. This picks out
communities of constitutions by methods of community detection and optimal modularity in which groups of constitutions which share  topics -- and thereby a directed edge -- in an amount above that expected by chance. Such a community constitutes a cluster \cite{newman}. 
Figure 2 displays the results of a network reconstruction of the full circuit along with two color codings of the network resulting from the application of two forms of clustering analysis to the network. The network is illustrated using spring embedding whereby densely connected nodes appear packed together. The network has the form of a ``constitutional caterpillar" with a temporal spine threaded through the network spanning 1789 to 2014 (Figure 2A). This temporal structure is very clear in the clustering coloring. Using community structure algorithms \cite{Girvan}  we observe (Figure 2B) three clear constitutional communities, each of which describes a span of time: epoch 1: from 1789 to 1936;  epoch 2: from 1937 to 1967; and epoch 3: from 1968 to 2014.  
Using a spectral technique for community detection we can further partition (Figure 2C) these network data into higher order communities \cite{newman}. This analysis maintains the chronological structure and illustrates the way in which clusters that are growing in absolute size (more constitutions in each) have evolved to encompass roughly decreasing ranges of time. 


Each constitution in the diffusion tree can be described in terms of its {\em transmission motif -- ``t-motif"}, a visualization of the indegree and outdegree for each constitution. A selection of these motifs is shown in Figure 3 with a full set in 
Supplementary Materials Figure S2.
The motifs demonstrate the variation to be found in balancing in-bound and out-bound influence for each constitution. Early constitutions tends to have few parents (e.g., Canada only has one -- the US constitution) whereas subsequent constitutions vary significantly in their ancestry. This variation can be explained thorough a combination of both time (earlier constitutions present more opportunities for imitation) and how representative, novel and applicable each constitutions is as a model for imitation.

\subsection*{Models for transmission}
We can gain further insights into the patterns of inheritance  by studying directly the distributions of indegree and outdegree across the entire dataset.  Figures 4A and 4B represent the pdf (probability density function) and cdf (cumulative distribution function) for the indegree for all constitutions. Illustrated in blue is the data and in orange the maximum likelihood parameter estimates for the best fitting distribution. The indegree distribution is well-captured by a Gaussian distribution with a mean of $8.8$ and a standard deviation of $2.9$. The estimated distribution does tend to slightly underestimate the mean but captures the tails very accurately. A straightforward interpretation is one of independent sampling of possible sources. The outdegree however, is quite different. Figures 4C and 4D show the best fitting Poisson distribution and the outdegree distribution. Whereas the mean is effectively recovered, the tails of the distribution are poorly fitted; the Poisson underestimates the number of constitutions with few offspring and overestimates the number of constitutions with many offspring. On the other hand consider Figures 4E and 4F where we show the best fitting negative binomial distribution to the data. This very accurately recovers the entire offspring distribution with maximum likelihood parameter estimates for the two shape parameters of the distribution as $r = 2.5$ and $p = .22)$  Recall that for a negative binomial $r$ describes the number of offspring observed  before no more offspring are generated and that the probability of producing an offspring is given by the value of $p$. We view this as a pure birth process as constitutions never die -- in the sense that they are always available as inspiration for a newly written constitution. Moreover, the negative binomial distributions are well known to be attractors of the Yule process \cite{karlin}, also known as ``preferential attachment" \cite{Hofstad} . The excellent fit of outdegree to this distribution has broader implications for connections between offspring number and longevity. In short, that we witness a small number of constitutions of relatively early constitutions of enduring influence.  All of this -- including the attendant modeling considerations -- is considered in some greater detail in the Discussion below.

\subsection*{Growth and Lifespans}
We are able to track the number of new constitutions written over time. We find statistical evidence for three epochs of authorship reflecting three distinct rates of growth (Figure 5 inset). These three growth phases coincide with the three temporal groupings of the transmission graph determined through spectral clustering. Hence there is an association between the growth rate and the detailed community structure of the graph. We also find significant variation in the lifespan of constitutions. The lifespan is defined as the first appearance to the last instance of influence.  There is a strong association between how early a constitution is written and how long it is observed to live. Unlike biological life spans nearly all constitutions ``die'' young (Figure 5).

\section*{Discussion}

We have searched for regular patterns of transmission in complex cultural artifacts. If there are cultural analogs to genotypes, and perhaps even phenotypes if we were to consider the broader context of constitutional influence, we should be able to observe their signatures in a temporally resolved study of evolving documents. Much like organisms that adapt to local environments, constitutions must be adapted to local cultural and legal conditions to be effective. And as with organisms, a great deal of variability in constitutions has been documented or inferred as derived from ancestral documents.

Our deeper discussion of the results starts with the labeling of the topics. We had an expert in constitutional law inspect the learned topics and provide labels for them corresponding to the dominant theme of the most probable words in each topic.  We note that providing labels for the learned topics is a challenging task due to the lack of ground truth. Assigning labels to topics in our setting is essentially projecting the learned topics onto one's conception of constitutional law and (admittedly) depends heavily on the individual involved contributing both bias and variance to the procedure. We assume that an expert in the field mitigates both of these effects and allows us to study the corpus using the learned topics.

Perhaps given the nature of the topic labeling problem (a general lack of ground truth) there is not much prior work on solving it. An early line of research examined whether commonly used predictive measures of topic models correlated with human interpretation of the topics and found that they did not \cite{Chang2009}. This previous work also was the first to use human experiments to evaluate the interpretation of learned topics. More recent work has focused on incorporating knowledge bases of topics (e.g., WordNet) directly into topic models in order to encourage the model to learn topics that are interpretable by biasing them to look like topics in the knowledge base \cite{Wood2016}. This is an interesting and difficult problem and further progress on it would enhance the results of this paper. 

The motifs (Figures 3) illustrate clearly how constitutions are ``cultural recombinants'' borrowing extensively from their ancestors. Constitutions vary in their hybridicity.
The motif variations suggest a constitution taxonomy, of {\em minor, major, idiosyncratic, and innovative} depending on where in the distribution matrix (divided via the median in both dimensions) the indegree and outdegree lie. As an example of a {\it minor} constitution, consider Switzerland 1848. It had no descendants and only two parents (Liberia 1847 and El Salvador 1843, both of which are probably explained by temporal proximity.) A {\it major} constitution, on the other hand, might be Thailand's 1932 Constitution, which established a constitutional monarchy and a European style administrative system: it had 15 parents and 33 offspring, making it the third most densely networked in the data.  {\it Idiosyncratic} constitutions include those of Burkina Faso 1991 and Lesotho 1983, with twelve and nine parents respectively, but only a single offspring each. Some 20\% of texts in the data have a child/parent ratio of $0.5$ or less, indicating more than twice as many parental relationships as offspring.  On the other hand, some 8\% of constitutions in the sample have a child/parent ratio of two or more, indicating relatively high levels of {\it innovation}.  Examples include Zambia's 1991 constitution, with 4 parents and 11 offspring, or Micronesia's constitution of 1990, with 8 parents and 24 offspring; in the latter case, it may be that the offspring are in fact those of the United States 1789, which was a very close model for Micronesian drafters.  In general, parent-child relationships are temporally proximate, and they are often geographically proximate.  This reflects the more general finding in the literature that time and space are powerful determinants of constitutional content.
This diversity highlights an important difference from biology where species of organisms show far less variation in the basic mechanics of transmission. 

Returning to the highlighted portion of Figure 1 to illustrate the mechanisms at play, consider Egypt's 1923 Constitution and its relationship with those of its descendants.  Examining the top ten topics in each text, Egypt 1923 shares multiple topics with Albania 1925 (topics act and public office)  and Iraq's 1925 documents (civil service and monarchy) Burundi (public office and labor) and one with Yugoslavia 1931 (mandate). No other constitutional dyad feature these combination of topics in the same density.  While the influence of Egypt's 1923 Constitution is well known to scholars of the Arab region, it also seems to share similarities with other documents drafted shortly thereafter in neighboring parts of Europe and Africa. This illustrates how our method can point scholars to look at new links that conventional analysis might not identify.

The most fecund constitution in our network is surprising at first glance: Paraguay's 1813 Constitution. 
 It makes sense, however, when one realizes that Latin America is the home to a plurality of constitutional texts, because it is a region of old nation states and frequent turnover \cite{ElGinsMel8}. Paraguay's was the first constitution adopted in Latin America after the Spanish Constitution of Cadiz of 1812. That document embodied an ill-fated attempt to establish a liberal constitutional monarchy in Spain, featuring equality under the law and popular sovereignty, and is recognized as a model for the constitutions of Norway of 1814, Portugal of 1822 and Mexico of 1824.  The top topic in this Constitution, ``language of law" consists of generic legal terms that are, of course, widely used in constitutional texts. So the influence was more formal than substantive.
 

Conversely, some canonical constitutions do not indicate the same kind of influence in our analysis that conventional analysis would expect.  For example, the 1936 Constitution of the Soviet Union is well known as a major step in the ideological development of communism in that it incorporated many rights that were never implemented. Yet at the level of ideas, much of this involved borrowing from extant models, such as the 1931 Republican constitution of Spain.  Perhaps unsurprisingly, there was little new that was in the USSR's constitution and so it has few children.  Similarly, the Weimar Constitution of 1919, which was thought to have embodied social democratic ideas \cite{Venter}, in fact was squarely within the topical mainstream of its time. With six parents and nine offspring, it is near the medians and its oldest direct ancestor is only 14 years prior to it. It shares three of its top ten topics ("geography'', "human rights'', and "education'') with SpainÕs Republican Constitution of 1931, which is regarded as an important and influential text. Its last direct descendent is the 1936 Constitution of the Soviet Union, with which it shares the topic "social development.'' This supports the claim that our method emphasizes ideological connections across text, because the Weimar Constitution is  generally considered to have been a structural model for France's 1958 Constitution \cite{Skach} though ideologically perhaps closer to that of the USSR.

The notion of ``cultural recombination" imports one kind of biological analogy to the evolution of constitutions. The distributions of the indegree and outdegree support different biological analogies. Consider again the striking result of the fit of the outdegree distribution to the negative binomial and the indegree to the Gaussian. A principled way to understand  these distributions is to derive them from suitable stochastic processes. The Gaussian distribution arises naturally from the sum of independent random variables with a well defined mean and variance.   Poisson distributions are attractors of the Galton-Watson process whereas negative binomial distributions are attractors of the Yule process (see e.g.\cite{karlin}). Both Poisson and negative binomial offspring distributions are observed frequently in biological systems. The Galton-Watson process was derived to explain the extinction of family names. The idea is that at each generation a parent can transmit their name to some number of $0,1,\dots, n$ offspring. Each parent samples the number of offspring independently from the same distribution. Our data support a negative binomial distribution so we shall focus on the Yule process. The Yule process is also well known as a preferential attachment process \cite{Hofstad} as it can be derived from an ``urn process"  in which balls of a given color are sampled in linear proportion to the number of balls already in each urn. The negative binomial distribution is derived by solving a simple recurrence equation describing the temporal evolution of a probability distribution of the form, $$P'_n(t) = -n\lambda P_n(t) +(n-1)\lambda P_{n-1}(t).$$ Here $P_n(t)$ is the probability of finding $n$ constitutions at time $t$. The rate of offspring production in some interval $\delta t$ is parameterized by $\lambda.$ Hence at a time $t$ a number $n$ of constitutions will decline through the addition of more offspring proportional to $n\lambda P_n(t)$ and increase through the production of offspring by the class $n-1$ at a rate $(n-1)\lambda P_{n-1}(t)$. If we establish an initial condition as the number of constitutions at the start of constitutional history as $1$, $P_0(0) = 1$, we find that, $$ P_n(t) = (\frac{n-1}{n-n_0}) e^{-\lambda n_0 t}(1-e^{-\lambda t})^{n-n_0}. $$ Which takes the  form of the negative binomial distribution in which we observe exactly $n_0$ offspring in $n$ trials with a success probability, $p = e^{-\lambda t}.$ For a formal exposition of preferential attachment dynamics illustrating the relationship of negative binomials to the special case of power laws see \cite{Ross}. 

We can test the assumptions of the Yule process by looking directly at the imitation dynamics of any given constitution. We simply plot the date on which the descendant of a given constitution was created against the order in which it was created. In Figure 6A we look at the evolution of the first $20$ constitutions. By far the majority have fewer than $10$ offspring and these offspring span a range of under $50$ years. However a few of these constitutions are exceptional. The most remarkable is the 1813 constitution of Paraguay that has provided material for $70$ descendant constitutions in a temporal range extending  $200$ years.  This is followed by the original constitution of the Unites States of America from $1789$ that produces 20 descendant constitutions, and over a span of $80$ years. The Canadian constitution of 1791 produces 11 descendants  over 150 years. Figure 6B includes the first $100$ constitutions, 6C the first  $200$ , and 6D all $591$ in the data set. A clear relationship between offspring number and longevity emerges consistent with preferential attachment in which a small number of constitutions are of dominant influence, these appeared early in constitutional history gaining a significant foothold, and with the vast majority of constitutions both short lived and producing less than 10 offspring.

The analysis of cultural recombination through a principled decomposition of textual artifacts suggests new domains of cultural inheritance. Unlike simple Mendelian systems, or simple learning models with homogeneous rules, we observe diverse patterns of variation in the way in which nations encode important moral and legal principles. Moreover we can obtain a principled definition of a  meme -- or unit of cultural transmission -- that goes beyond the single ``word" and captures highly linked sets of words expressing a functional, legal category -- much the way a gene, composed of linked sets of nucleotides -- contributes to a function.  Nations differ in their debt to the past and their original contributions to the future. This allows us to speak in a rigorous fashion about phylogenetic concepts like analogy and homology when it comes to a cultural artifact. This has been an area of active research which includes the formal analysis of cultural and symbolic systems \cite{sforza, Nowak1, Nowak2}, experimental approaches to cultural transmission \cite{Henrich,Mesoudi2}, and qualitative frameworks of integration \cite{Mesoudi}. At this point in time the status of key phylogenetic concepts applied to culture is in flux \cite{mace},  we favor an instrumental approach defining cultural analogy and homology strictly in phylogenetic terms. 

We  suggest that the ``semantic" interpretation of a given constitution and its practical legal impact is what we  mean by the phenotype. We might expect many different genotypes to be neutral in that their interpretations are equivalent, and that constitutions vary in their ``penetrance", that is their influence on cultural practices. 

This approach builds on prior research related to concepts such as ``citation backbones" \cite{Gualdi} in which citations to prior publications form a tree-like structure from which novel papers descend,  patent backbones in the automobile industry \cite{Lin}, skewed patterns of borrowing in human designed artifacts \cite{Eldredge}, patterns of word borrowing \cite{Nelson-Sathi} and the evolution of programming languages \cite{Valverde}.

Reconciling statistical patterns of influence with potential biases and patterns in thinking and writing will bring us closer to frameworks that connect methods of mathematical science with objects of psychological and humanistic interest in the service of new models and theories of cultural transmission and influence. The evolution of the law with its rich textual and interpretive traditions provides a nearly ideal model system.

\pagebreak
\clearpage

\begin{figure}[ht]
\centering
\includegraphics[scale=1.6]{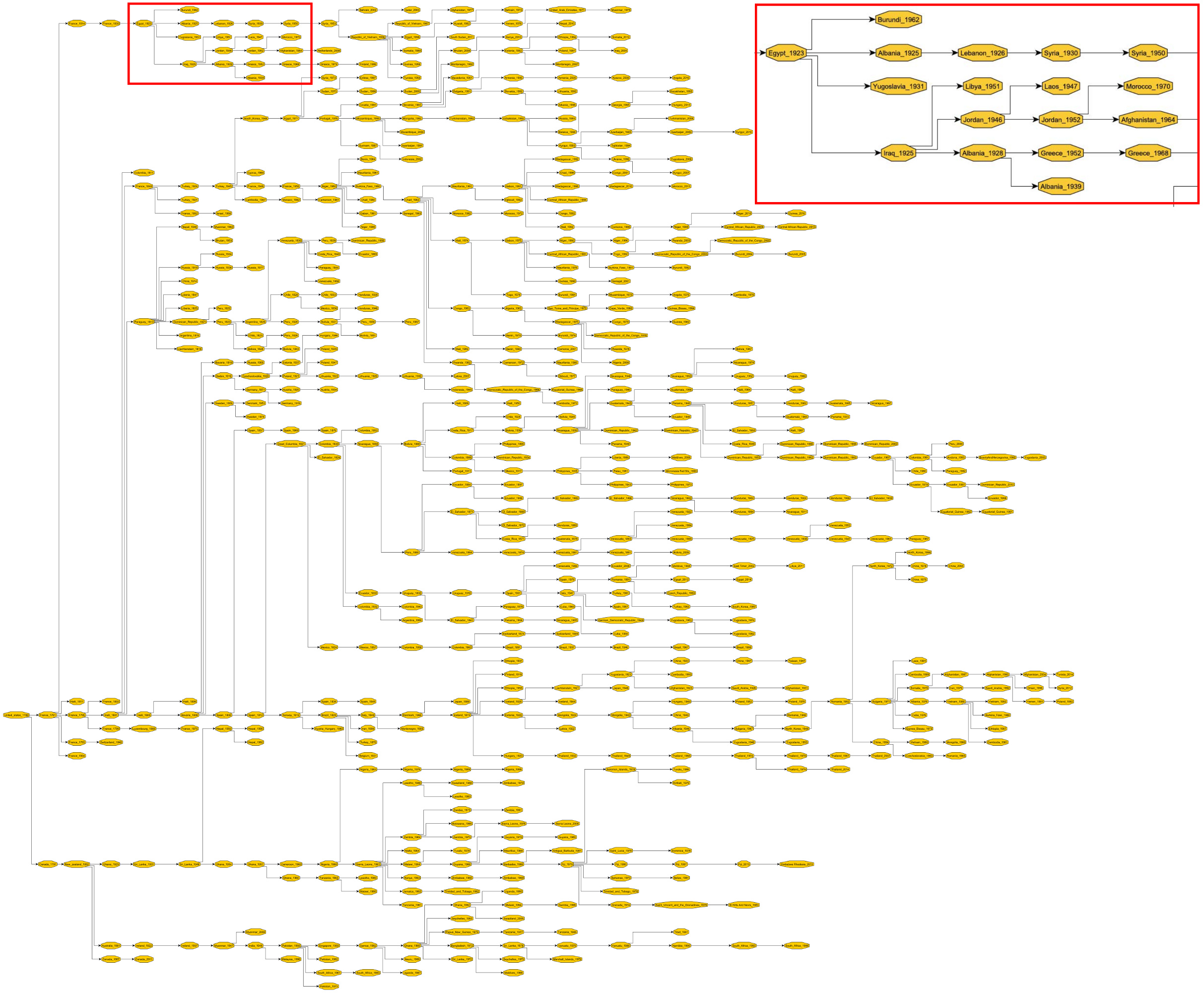}
\caption{The ``family tree" of constitutions. The United States constitution of 1789 is the root and thus the ``Last Universal Common Ancestor Constitution". Any other constitutions is deemed as having as its most recent ancestor the closest earlier constitutions where distance is measured as the KL-divergence of the former to the latter. The size of this tree makes it difficult to render so that the constitution country and date are legible. A detail of the tree around the Egyptian constitution of 1923 is provided in the upper inset. Note the fertility of that constitution, as well as the sterility of the constitutions of Burundi (1962), Morocco (1970), and Albania (1939). The last of these is particularly interesting as we see a line of descendants issuing forth from the Albania constitution of 1925. The Albanian Constitution of 1939 was an imposed, fascist document that drew on earlier models, but had little purchase after World War II. The most frequent topic, ``subnational government" is found in such proportions in only one other, earlier text. So, earlier versions of constitutions can have patterns of transmission that do not include all of their descendants.A pdf document of this tree, easily magnifiable, can be found at https://www.math.dartmouth.edu/$\sim$rockmore/kl-tree.pdf. }
\end{figure}

\clearpage

\pagebreak

\pagebreak
\clearpage
\begin{figure*}[ht]
\centering
\includegraphics[scale=0.12]{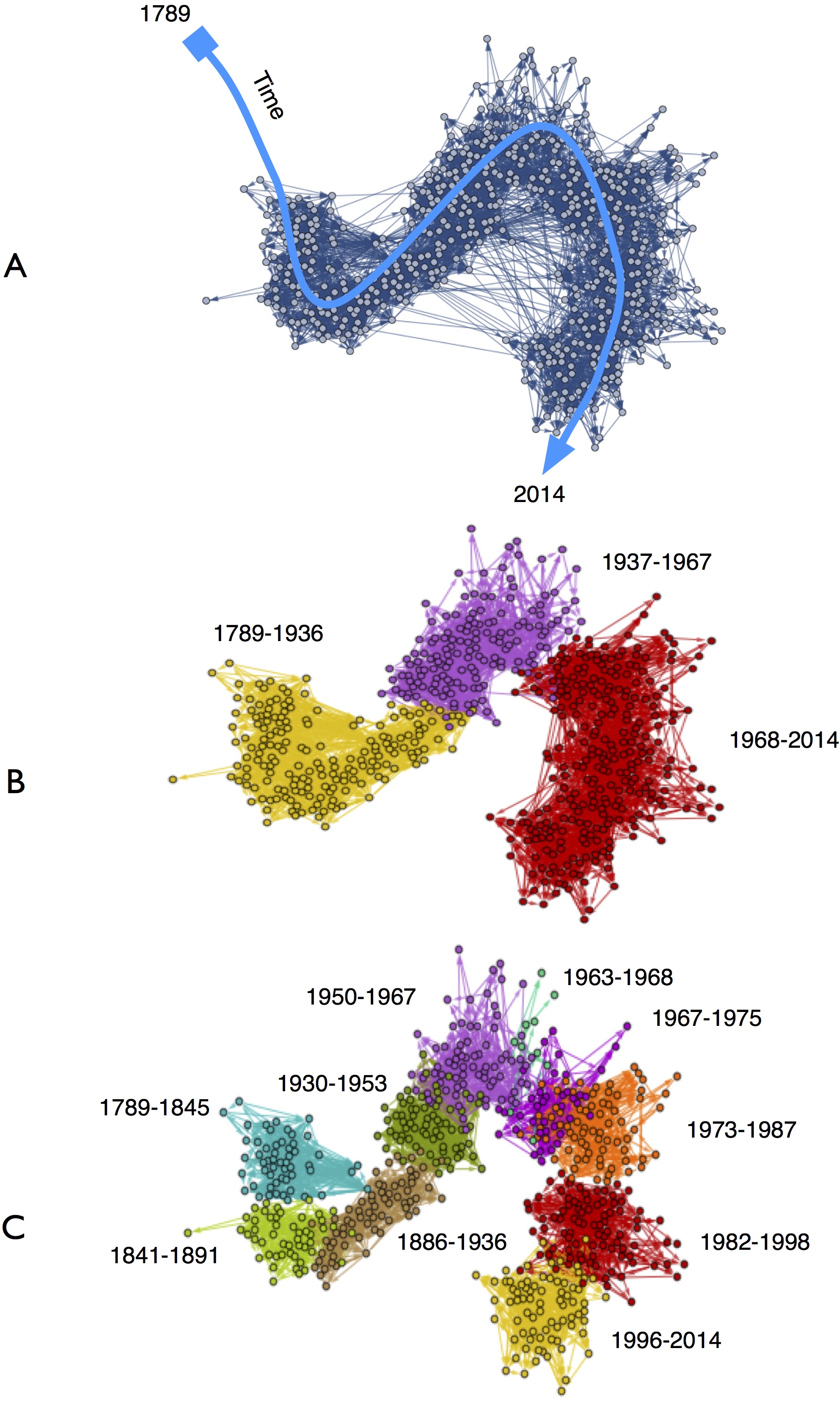}
\caption{(A) Spring embedded reconstruction of constitutional diffusion network. Nodes correspond to constitutions and directed edges encode topic borrowing. The blue arrow traces time forward through the network starting in 1789 and ending in 2014.  Time is the dominant factor in explaining the geometric form of the network.  (B)  Application of a community detection algorithm  to  the thresholded diffusion tree reveals three clear epochs of constitutional  inheritance. The oldest epoch spans $147$ years and contains $175$ constitutions generating an average of $1.2$ constitutions per year. The second epoch spans $30$ years and contains $148$ constitutions and an average of $4.9$ constitutions per year.  The third epoch spans $46$ years and contains $267$ constitutions and an average of $5.8$ constitutions per year. The rate at which constitutions are being written has increased through time whereas the temporal influence of constitutions into the future has contracted. (C) Use of more sensitive optimal modularity methods provides a decomposition of each of these epochs into a further three epochs.   Each induced cluster preserves the largely temporally contiguous ordering demonstrating that time remains a dominant dimension of variation  at the microscopic level. }
\end{figure*}

\pagebreak
\clearpage

\begin{figure*}[ht]
\centering
\includegraphics[scale=0.13]{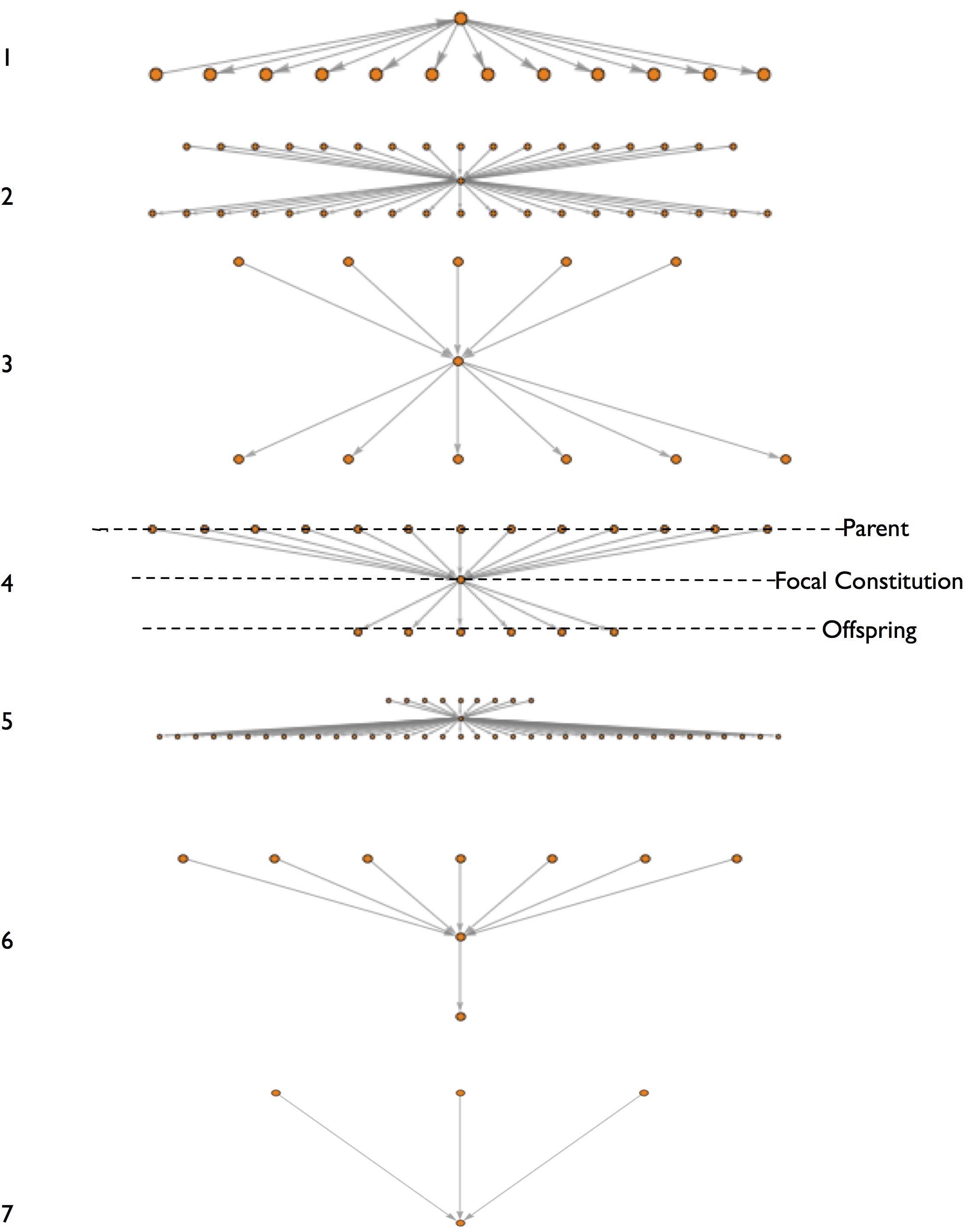}
\caption{Each constitution in the diffusion tree can be described in terms of a {\em transmission motif}, which visualizes the indegree and outdegree for each target constitution.  The motifs demonstrate the balance between in-bound and out-bound influence for each constitution in terms of a threshold number of topics that are borrowed. (1) Early constitutions tends to have few parents, e.g.,  Canada (1791)  only has one (the U.S.(1789)  constitution, the leftmost node in Figure 1). Subsequent constitutions vary significantly in their ancestry: (2) Iceland (1874)'s constitution has many parents and many offspring; (3) Bolivia (1826) constitution has fewer parents and few offspring (4) Venezuela (1830) exhibits many parents and few offspring; (5) South Korea (1948) has few parents and many offspring;  (6) Albania (1976) has several parents and only one offspring; (7)  Montenegro (1992) has no offspring. This variation in parentage and fertility can be explained thorough a combination of both the time at which they were written and the tendency to preferentially attach to a small number  of highly favored models for imitation.   }
\end{figure*} 

\clearpage

\begin{figure*}[ht]
\centering
\includegraphics[scale=0.15]{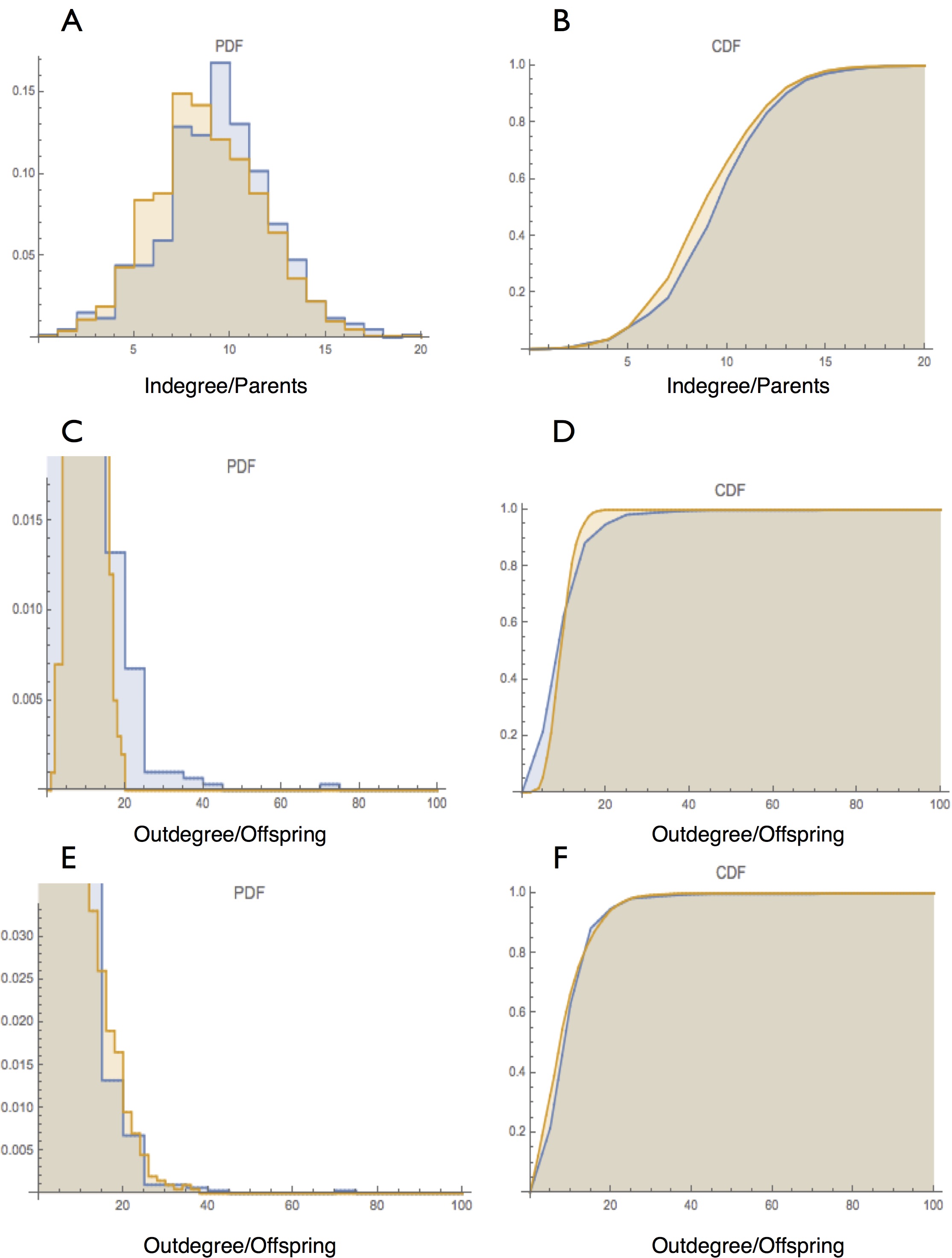}
\caption{Illustrated in blue are the inferred connectivity data and in orange the maximum likelihood parameter estimates for the best fitting distributions for constitutional indegree (A,B) and outdegree (C,D,E,F). The indegree is best approximated by a Gaussian distribution with a mean of 8.8 and a standard deviation of 2.9. Figures 4C and 4D plot the outdegree distributions and the best fitting Poisson distribution. Whereas the mean is effectively recovered, the tails of the distribution are  poorly fitted.  The Poisson underestimates the number of constitutions with few offspring and overestimates the number of constitutions with many offspring. In Figure 4E and 4F we show the best fitting negative binomial distribution to the data. This very accurately recovers the entire offspring distribution with maximum likelihood parameter estimates for the two shape parameters of the distribution as $r = 2.5$ and $p = .22$.    }
\end{figure*} 
\clearpage

\pagebreak
\clearpage

\begin{figure*}[ht]
\centering
\includegraphics[scale=0.14]{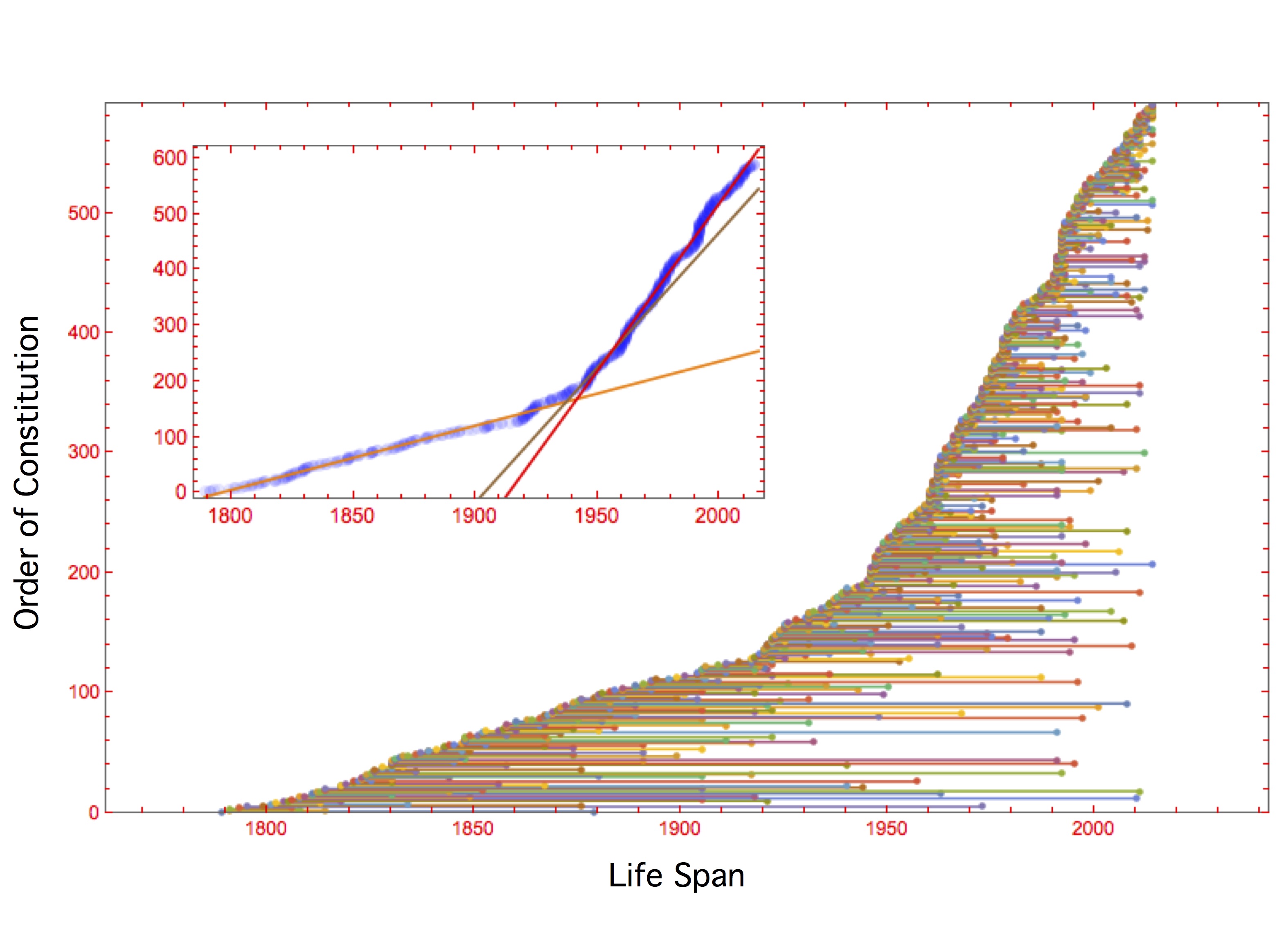}
\caption{Growth and life span of constitutions. The inset figure superimposes the $n$ best-fitting piece-wise linear regressions over the growth rate of constitutions (show in the larger image). We discover that $n=3$ and that these three growth rates correspond to the three epochs uncovered through the community structure analysis. We also show the life span of constitutions (first appearance to last recorded influence), with the life-span plotted against the order of appearance of a constitution in the corpus.  We clearly see how the earliest constitutions exert the longest influence on descendant constitutions -- a result strongly in accord with the findings supporting a form of preferential attachment rule of influence.   }
\end{figure*}

\clearpage

\pagebreak
\clearpage
\begin{figure*}[ht]
\centering
\includegraphics[scale=0.15]{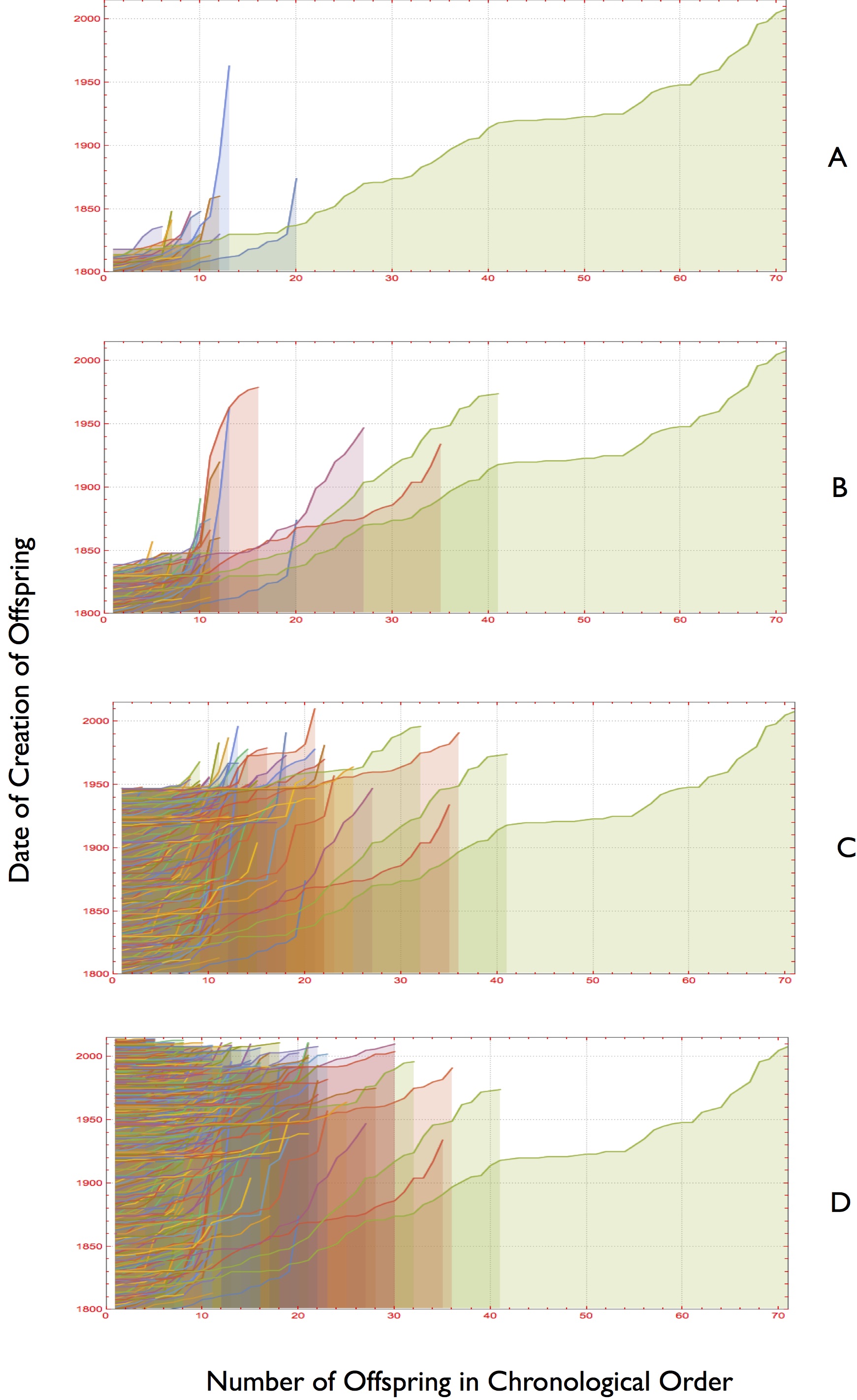}
\caption{Fecundity and influence of constitutions.  On the $x$-axis are the number of descendant constitutions arranged in chronological order and on the $y $-axis the date of their appearance. In Panel A we plot the first 20 constitutions. In Panel B the first 100. Panel C the first 200. Panel D all 591.  Most constitutions have few descendants and these appear over a relatively short span of time. Constitutions with many descendants tend to span longer periods of time. Most of the longest-lived constitutions in terms of influence/borrowing were written in the first of the three epochs of constitutional history (as in Figure 4A).  }
\end{figure*}
\clearpage
\clearpage

\newpage

\section*{Supplementary Materials for The Cultural Evolution of National Constitutions: Supporting Information}

\section*{Materials and Methods} 

As explained, our results and methodology depend on the use of topic models (see e.g.,  \cite{Blei-ACM}) and diffusion networks. Topic models are statistical models to learn the underlying structure of a corpus of documents. There are many flavors of topic model. We use the Latent Dirichlet Allocation (LDA)~\cite{BleiNJ03} probabilistic generative topic model. The underlying topics are represented as latent variables in a hierarchical Bayesian model. A generative model is assumed to be responsible for the observed documents and the word distributions of each topic. The topic proportions of each document can be learned via estimation of the posterior distribution of latent variables conditioned on the observed documents. The topic representations of the constitutions then form the underlying data for the inference of the diffusion network a la \cite{tkdd/Gomez-RodriguezLK12}. Some details of this now follow. 

\subsection*{Materials}
Our basic materials are 591 constitutions in English obtained from the publicly available and accessible Comparative Constitutions Project website (http://comparativeconstitutionsproject.org/). A complete list of the constitutions we use is in Table S1. 

\subsection*{Methods}

\subsubsection*{Topic modeling}
The foundation of our text analysis is the use of a  form of {\em topic modeling} on the corpus of 591 constitutions from which we derive a diffusion network for the inferred topics. A {\em topic} is a probability distribution over a fixed {\em vocabulary} derived from a text {\em corpus}.  The corpus is composed of {\em documents}, where a document consists of a set of (possibly non-unique) {\em words} from the vocabulary. 

We obtained PDF versions of the constitutions from \cite{ConstProj} and converted them to text files.  Table S1 provides a list of the constitutions. The documents in the corpus are contiguous blocks of text extracted from partitioning the constitutional texts.  We set a document length of 500 words and also require that documents respect the borders of constitutions (i.e., no document straddles multiple constitutions). If the length of a document is too long then the learned topics will put similar probability on many words and thus will not capture our intuitive notion of a topic. If the document length is too short the resulting topics are overfit to specific documents as there is insufficient data to learn general topics that can be used across the corpus. In addition, the choice of document length depends on the type of structure we are interested in, short document lengths are good for learning localized topics that are specific whereas longer document lengths learn smooth topics that explain large portions of the corpus. 

We use a standard methodology for further preprocessing the documents by stemming the documents using the well known NLTK stemming package (http://www.nltk.org/api/nltk.stem.html), removing English stopwords as well as words that appear less than 20 times across the entire corpus. We also remove words that appear in over 90\% of the corpus. The resulting vocabulary consists of 3,546 unique terms.   We then computed the number of occurrences of each word in the vocabulary in each document so that each document is represented by a 3,546 dimensional vector where the $i$th entry contains the number of occurrences of token $i$ in the document. This is a {\em bag of words} representation. (i.e., that the order of words does not matter, also referred to as {\em exchangeable}) and additionally we assume that the order of the documents, both within and between constitutions, does not matter.

We then topic model the document corpus using  the Latent Dirichlet Allocation (LDA)~\cite{BleiNJ03} probabilistic generative topic model. 
In the topic model a document is viewed as a mixture of topics where the underlying topics are represented as latent variables in a hierarchical Bayesian model. A generative model is assumed to be responsible for the observed documents and the word distributions of each topic. The topic proportions of each document through the posterior distribution of the latent variables conditioned on the observed documents.

To set notation let our corpus $\mathcal{D}$ be defined as a set of $N$ documents, $\mathcal{D} =\{d_1,d_2,\ldots,d_N\}$.  Let  $l$ denote the document length and break a constitution into multiple documents, respecting constitution boundaries. A \textit{topic} is a distribution over a fixed vocabulary $V$ and can thus be represented by a vector $\beta \in \mathbb{R}^m$, $\beta_i \geq 0, \sum_i \beta_i = 1$, where  $m$ is the size of the vocabulary and the $i^{th}$ entry is the probability of picking word $i$ from this topic. We denote the proportion of document $d_i$ that is made up of topic $i$ by $\theta_i \in \mathbb{R}^{K}$, where $K$ is the number of topics, where $\theta_i \geq 0$ and $\sum_i \theta_i = 1$.
Given the $\ell$th word in document $i$, $w_{i\ell}$,  let $z_{i\ell}$ indicate which topic the token $w_{d\ell}$ is drawn. Let $Dir(\eta)$ denote the Dirichlet distribution with parameter $\eta$ and $Mult(\theta_i)$ denote the multinomial distribution over the distribution $\theta_i$. 


The specific generative process underlying LDA is as follows:
\begin{enumerate}
  \item Fix $K$, the number of topics
  \item For each topic $k$, draw $\beta_k \sim Dir(\eta)$
  \item For each document $d_i$:
  	\begin{itemize}
  		\item Choose topic proportions $\theta_{d_i} \sim Dir(\alpha)$
  		\item For each word position $w_{d_i\ell}$:
  			\begin{itemize}
  			\item[--] Choose a topic indicator $z_{d_i\ell} \sim Mult(\theta_{d_i})$
  			\item[--] Choose a word $w_{d_i\ell} \sim Mult(\beta_{z_{d_i}})$
  			\end{itemize}
  	\end{itemize}
\end{enumerate}

Note that LDA depends on four parameters, the Dirichlet parameters $\alpha,\eta$, the number of topics $K$ and the document length $l=500$. In order to expedite the mixing  of  the Markov chain and reduce experiment time, we fix $\alpha$, $\eta$ and $l$, and vary the value of $K$. Choosing an appropriate number of topics for a given corpus is a problem of \textit{model selection}. We carried out 5-fold cross-validation to optimize $K$. Specifically, we split the corpus evenly into 5 folds which are used to define training and testing sets to evaluate parameter configurations.  For each configuration of $K$ we hold out one of the folds as a test set, $W^{\mathrm{test}}$, and use the other four as the training set, $W^{\mathrm{train}}$.  We ran the Gibbs sampler for LDA on the training set for $10,000$ iterations which produced samples from the posterior distribution which were then used to evaluate the likelihood of the test set, $p(W^\mathrm{test}|K)$ which measures the generalization ability of the model.  Unfortunately, the computation of the held-out likelihood, $p(W^\mathrm{test}|K)$ is intractable so we adopted the Chib-style estimation in~\cite{WallachMSM09} to efficiently approximate it.  The values of $K$ and $l$ that obtain the highest overall held-out likelihood over the five folds are chosen for the rest of our analysis.   Figure S3 shows the effect of varying $K$ from which we see the optimal value is $K=100$.

\subsubsection*{Inferring diffusion networks}
The topics, $\beta_{1:K}$, that are learned with LDA represent high level ideas and each constitution $C_i$ can be represented by the proportion of topics it exhibits, $\theta_i$ (which we described how to compute above). As demonstrated in experiments, these discovered topics correspond to high level legal aspects, such as human rights, international agreements, and economic systems. By treating a topic as the unit in a diffusion and tracking the occurrence of each topic at each constitution over time, we can learn an underlying \textit{diffusion network} by which topics spread through constitutions over time, thus uncovering the \textit{diffusion patterns} of legal evolution over time.

We follow the method of Rodriguez et al.~\cite{tkdd/Gomez-RodriguezLK12} for inferring diffusion networks. We define a \textit{cascade} $c$ as a set of pairs $(i,t_i)$, indicating that cascade $c$ was observed at node/constitution $i$ at time $t$. Each topic, $\beta_k$, will have an associated cascade, $c_k$, so that $(i,t_i) \in c_k$ means that topic $k$ has spread to constitution $i$ at time $t$. To determine if a topic has spread to a constitution, we set a threshold $\tau$ and we say the topic is observed at the nodes/constitutions whose proportion of this topic are among the top $\tau$ percent 
When a topic does not spread to a constitution, $C_j$, we set $t_j = \infty$.
Note that we do not observe the path by which topics are spreading but only where topics have spread to at a given time.

Having defined cascades, we describe a probabilistic model of how they diffuse through constitutions. Specifically, we denote the probability that a cascade $c$ is transmitted from node $i$ to node $j$ as $P_c(i,j)$, where $t_j \geq t_i$, indicating that a constitution can only be influenced by its predecessors. In our experiment, we take $P_c(i,j)=e^\frac{-(t_j-t_i)}{\hat{\alpha}}$, where $\hat{\alpha}$ is the \textit{diffusion parameter} and $t_i$ and $t_j$ are the timestamps of constitutions $i$ and $j$.

A diffusion network $G$ is a \textit{directed graph} \cite{clrs} where an edge from node $i$ to node $j$ indicates that topics can diffuse from constitution $i$ to constitution $j$.  We note that any directed graph can be represented as the union of the set of \textit{spanning trees} \cite{clrs}, i.e.\ sub-graphs that connect all of the nodes and that have no cycles. The inference process produces an optimal network able to accommodate the observed cascades. To get a sense of what the process produces, see Figure S1 for an inferred diffusion network derived from the methodology discussed here on a subset of $99$ constitutions. 

First, we define the probability that a cascade $c$ is consistent with a given tree structure $T$ (where the edges in $T$ obey the ordering of time stamps) to be the following:
\begin{equation}
\label{en:c_given_T}
P(c|T) \propto \prod_{(i,j)\in T}P_c(i,j)
\end{equation}
Notice that Eq.1 assumes that all edges in $T$ are independent but that $P_c(i,j)$ are conditional probabilities so that Eq.~\ref{en:c_given_T} defines a Markov process.  Notice that Eq.~\eqref{en:c_given_T} only depends on the edges in the tree as nodes not observed in a cascade have and infinite time of observation\footnote{In~\cite{tkdd/Gomez-RodriguezLK12} the probability of a cascade spreading from a given node or dying off at the node is modeled with a Bernoulli random variable in order to account for the fact that cascades usually do not reach all nodes and thus controlling the size of the cascades.  However, the probability of spreading and not spreading turns out to be a constant in the optimization used to infer a diffusion network so we ignore it here and it turns out to be computationally advantageous to control the complexity of the inferred diffusion network using a constraint on the number of edges in the inferred diffusion network.}.

Using Eq.1 we define the probability of observing a cascade given an arbitrary diffusion network $G$ as:
\begin{equation}
P(c|G)=\sum_{T\in \mathcal{T}(G)} P(c|T)p(T|G)
\end{equation}
where $\mathcal{T}(G)$ is the set of all spanning trees of $G$ and we assume $p(T|G) = 1 / |\mathcal{T}(G)|$ is uniform over all spanning trees $T \in \mathcal{T}(G)$. Lastly, we define the probability of observing all cascades, $C = \{c_k\}_{k=1}^K$, one for each topic, for a given diffusion network $G$ as:
\begin{equation}
P(C|G)=\prod_{c \in C}P(c|G).
\end{equation}
The goal is then to find the maximum likelihood diffusion network by maximizing Eq.3 with respect to $G$ over all possible directed graphs with consistent time stamps (directed edges only emanate from earlier constitutions and terminate in later constitutions). Formally, we need to solve the following optimization problem:
\begin{equation}
\hat{G}=\mbox{argmax}_{|G| \leq k} P(C|G)
\end{equation}
where the constraint, $|G| \leq k$, indicates that the number of edges in $\hat{G}$ be less than $k$.  This constraint provides complexity control since the graph consisting of all edges from a constitution to later constitutions is a trivial solution and because as mentioned above cascades usually only consist of a subset of constitutions.  Optimizing Eq.~4 is NP-hard, however, an efficient greedy algorithm that obtains a near-optimal solution due to the submodularity of the problem~\cite{tkdd/Gomez-RodriguezLK12}.  In addition, we use a heuristic that stops the algorithm from adding new edges (and thus terminating) when the objective function in Eq.~4 reaches $80-90\%$ of an upper bound derived in~~\cite{tkdd/Gomez-RodriguezLK12}.  This allows us to avoid using expensive cross-validation when setting the complexity, $k$, of the model.

A key parameter is the threshold $\tau$, which sets the fraction of topics viewed as important at a given constitution.  In order to set $\tau$, we varied $\tau$ between $0$ to $0.8$, and inferred the diffusion network for each of the values. We optimize $\tau$ relative to the parameters of the mean in- and out-degrees of all inferred diffusion networks sat each parameter. This can be found in Figure S4. Note the hump shape of the means in increasing $\tau$, reflecting the gradual accumulation of possible paths in increasing $\tau$ and then a tailing off as the optimization aspect of the diffusion network construction begins to winnow edges.  
After observing both in- and out-degrees reach peaks at $0.3$, we investigate the robustness around $0.3$. For a set of thresholds, most densely sampled around $0.3$ we create a vector of in-degrees ordered according to the year of the constitution and a vector of out-degrees similarly ordered. Then for the in-degrees a (symmetric) matrix is constructed computing the Pearson correlation of the $i,j$ entry, and similarly for the out-degrees. Figures S4 and S5 show the heat map of values. The farther away you are from the diagonal, the farther you are in threshold difference (most densely sampled near 0.3). The slow roll-off in color reflects the robustness of the calculation in that region -- i.e., the in-degree and out-degree orderings are not changing much as threshold varies between $0.2$ and $0.4$. All of this motivates a choice of $\tau = 0.3$. 

The final diffusion tree produces for each constitutions a set of direct descendants and ancestors, thereby giving rise to the indegree/outdegree ``motifs". A full list of motifs is presented in Figure S2. The full tree can be found at www.math.dartmouth.edu/$\sim$rockmore/FullConstDiffNet.pdf. A detail of the full tree is given in Figure S5. 


\begin{figure*}[h]
\centering
\includegraphics[scale=.3]{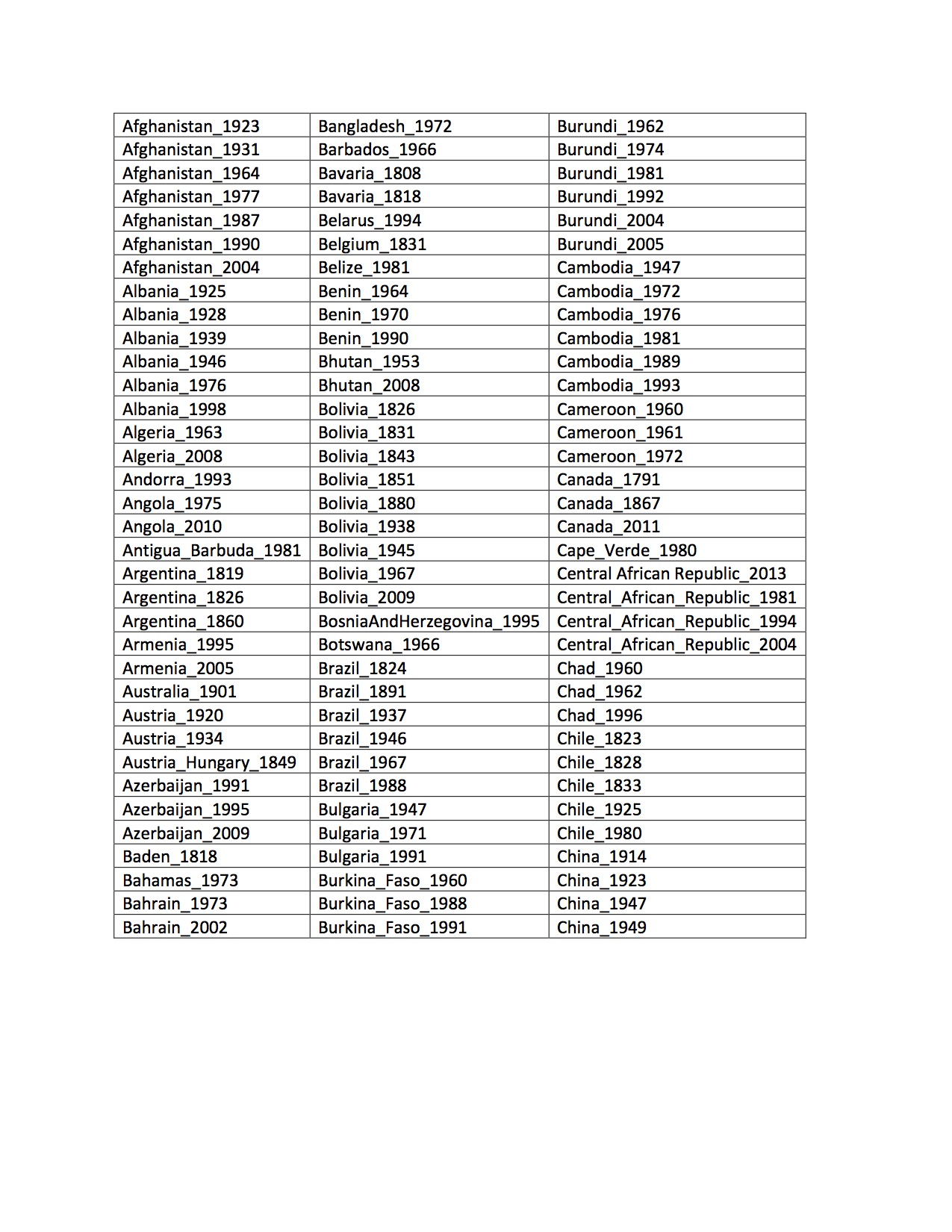}
\caption*{{\bf Table S1(a).} List of constitutions in our corpus in alphabetical order -- Table 1 of 5.}
\end{figure*}
\begin{figure*}[h]
\centering
\includegraphics[scale=.3]{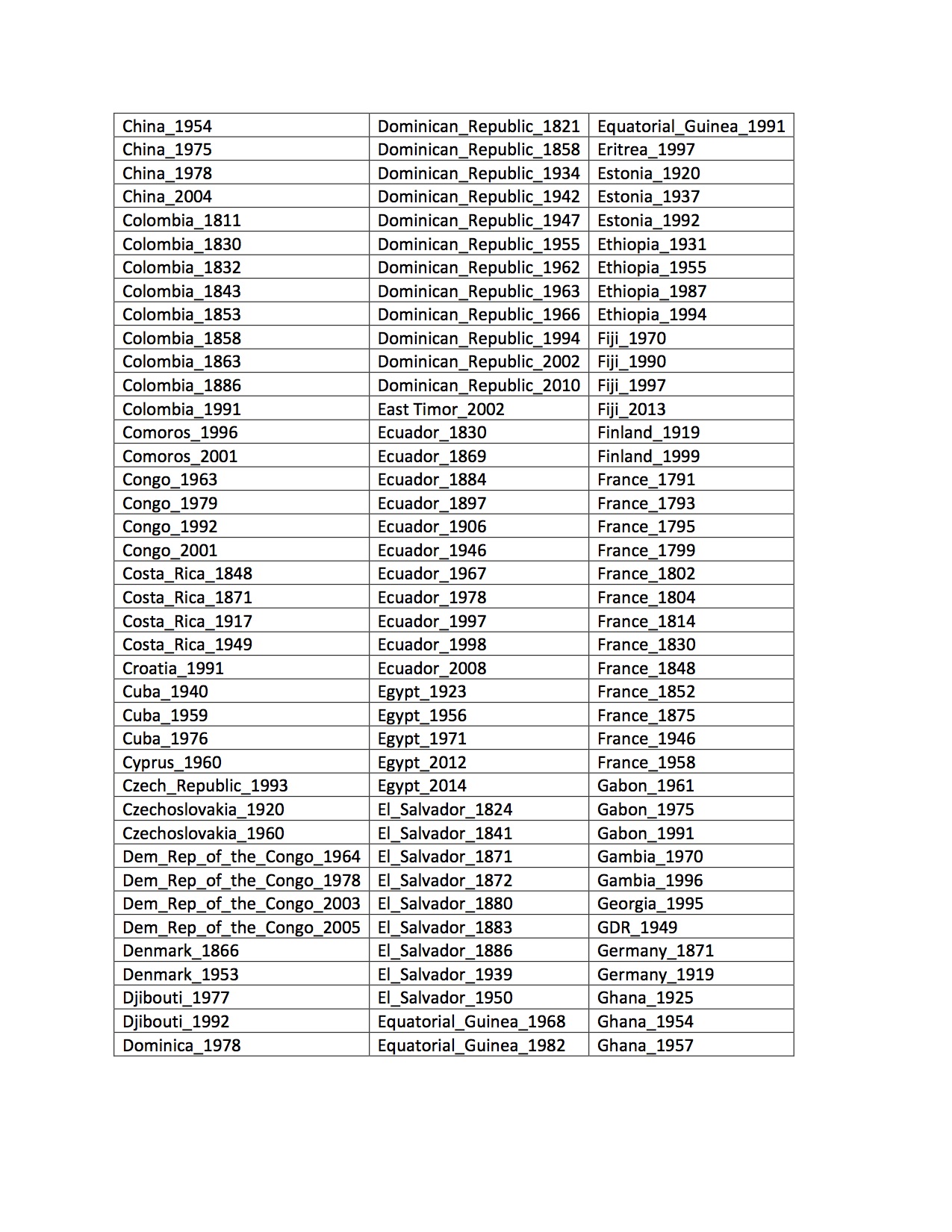}
\caption*{{\bf  Table S1(b).} List of constitutions in our corpus in alphabetical order -- Table 2 of 5.}
\end{figure*}
\begin{figure*}[h]
\centering
\includegraphics[scale=.3]{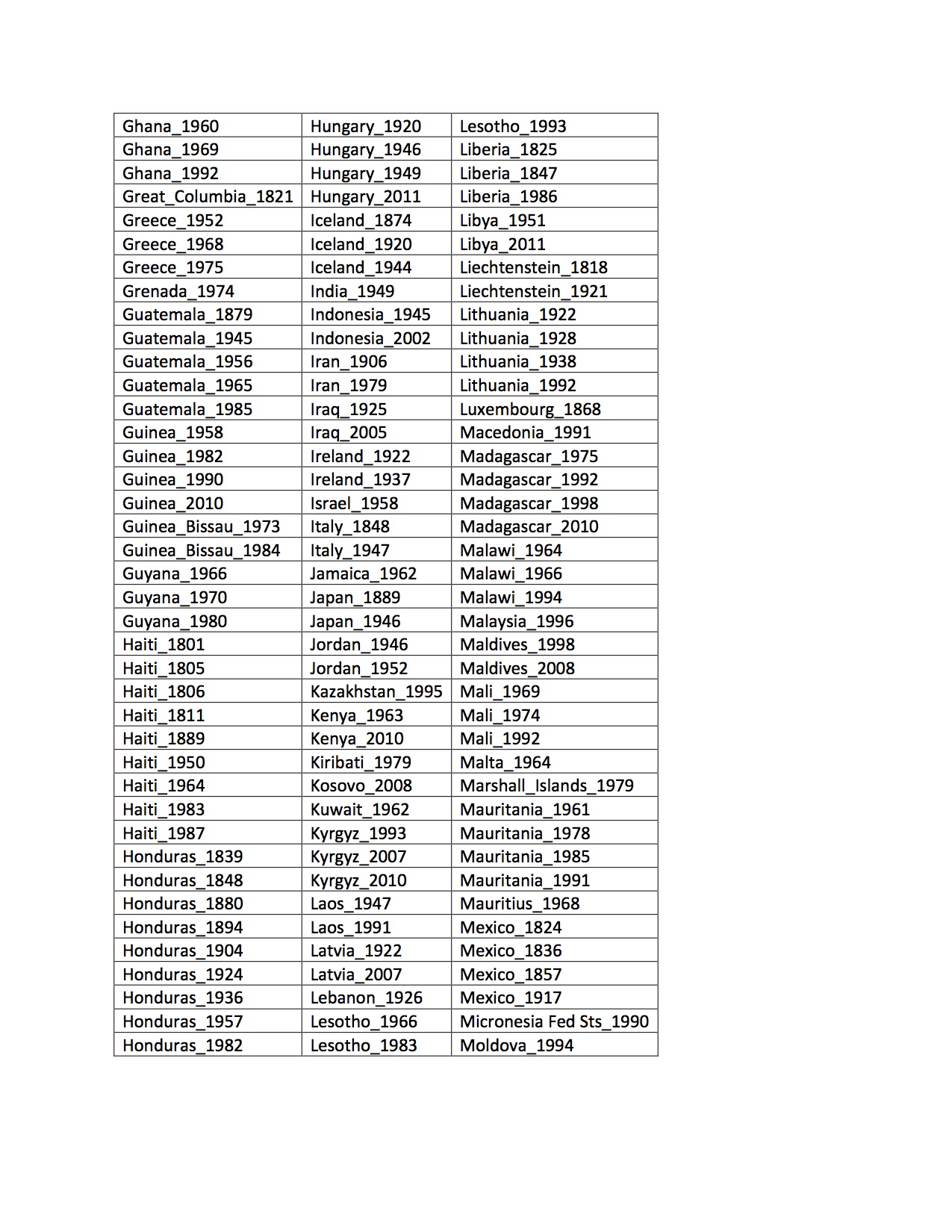}
\caption*{{\bf  Table S1(c).} List of constitutions in our corpus in alphabetical order -- Table 3 of 5.}
\end{figure*}
\begin{figure*}[h]
\centering
\includegraphics[scale=.3]{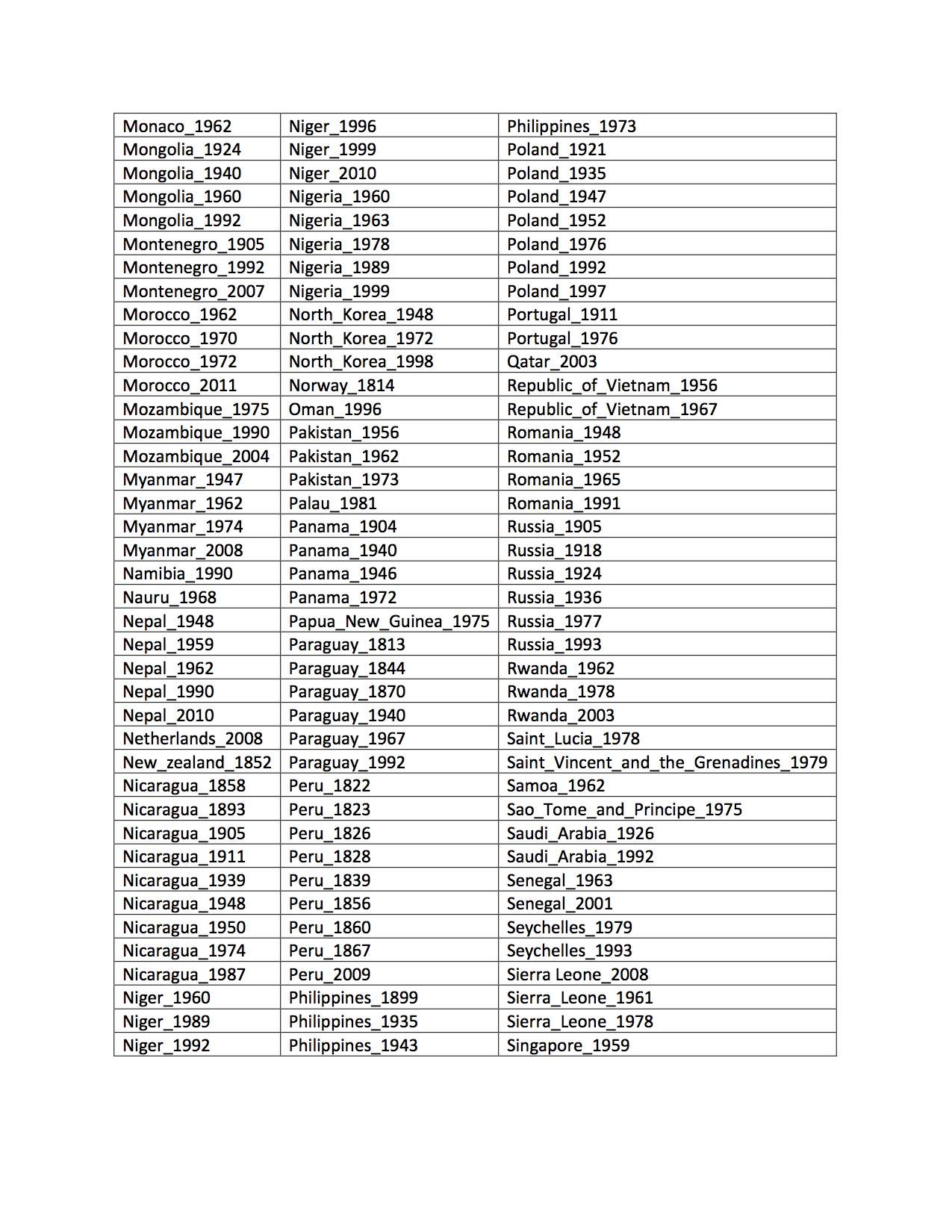}
\caption*{{\bf  Table S1(d).} List of constitutions in our corpus in alphabetical order -- Table 4 of 5.}
\end{figure*}
\begin{figure*}[h]
\centering
\includegraphics[scale=.3]{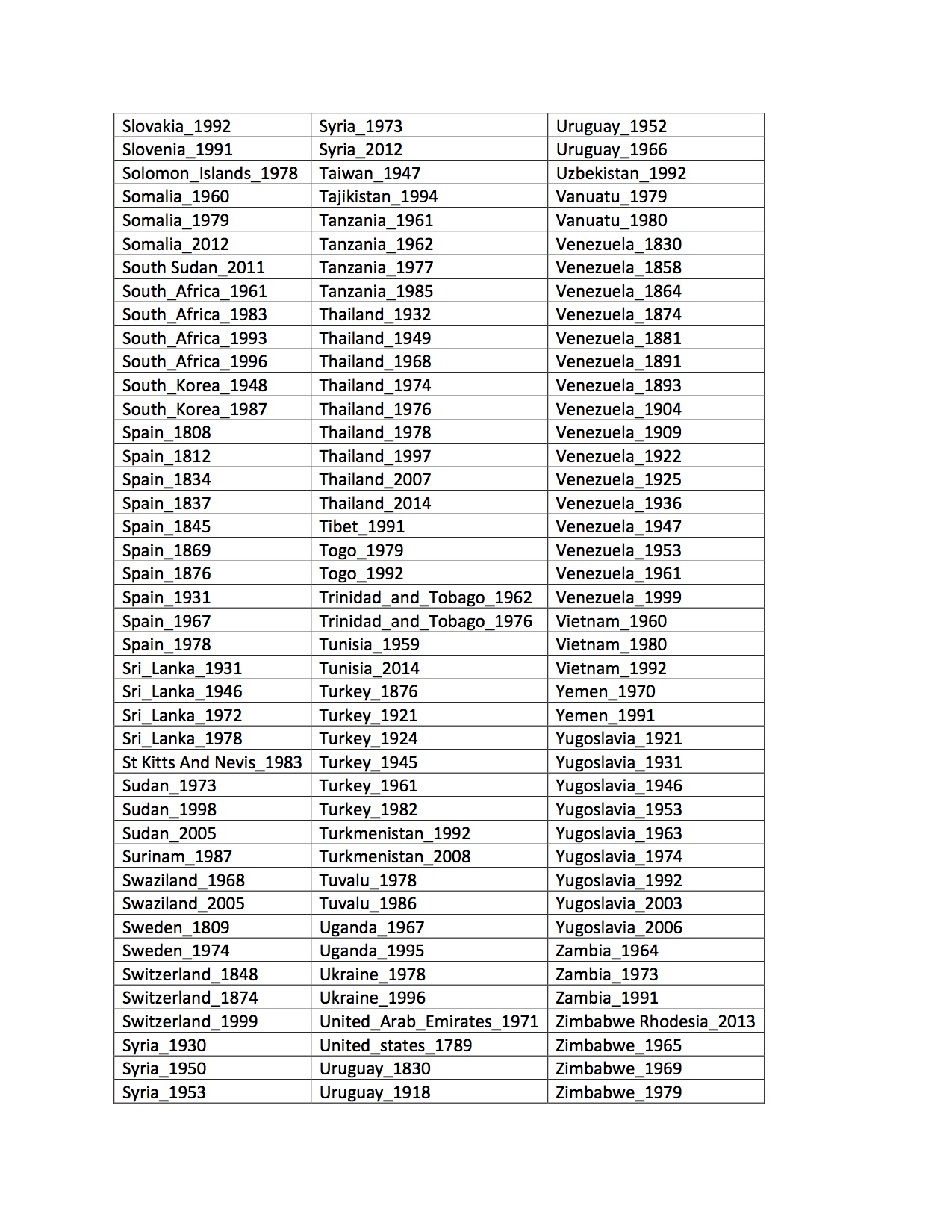}
\caption*{{\bf   Table S1(e).} List of constitutions in our corpus in alphabetical order -- Table 5 of 5.}
\end{figure*}

\pagebreak

\begin{figure*}
\centering
\includegraphics[scale=.2]{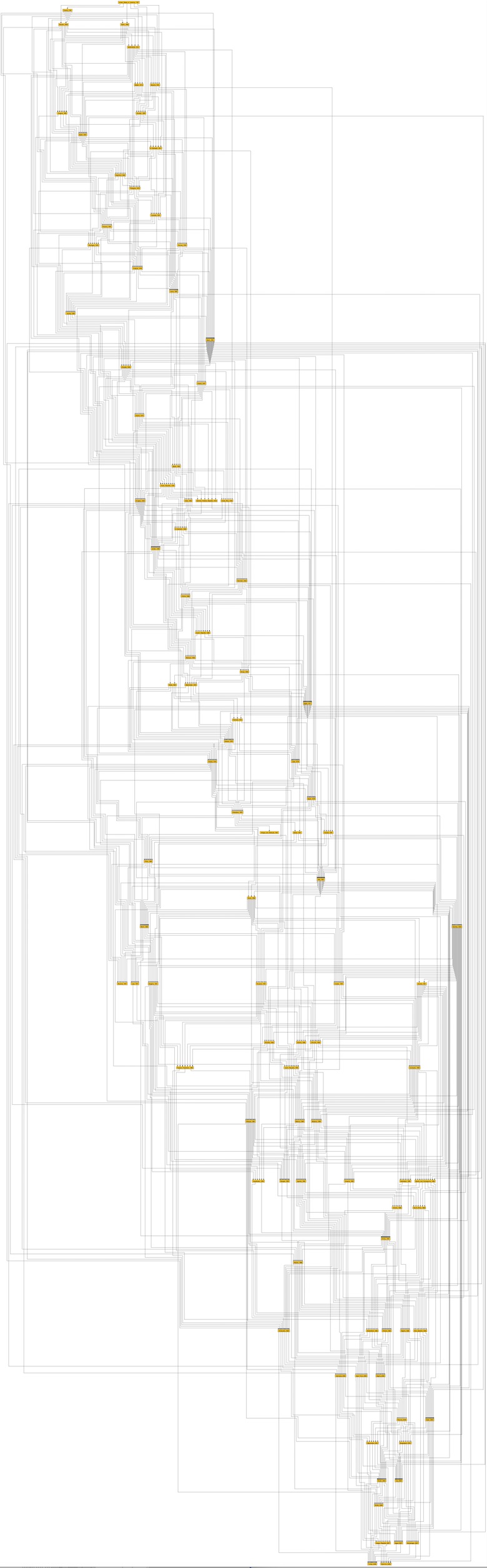}
\caption*{{\bf Figure S1.} Learned diffusion network on a subset of ninety-nine constitutions.}
\end{figure*}

\begin{figure*}
\centering
\includegraphics[scale=.25,angle=90,origin=c]{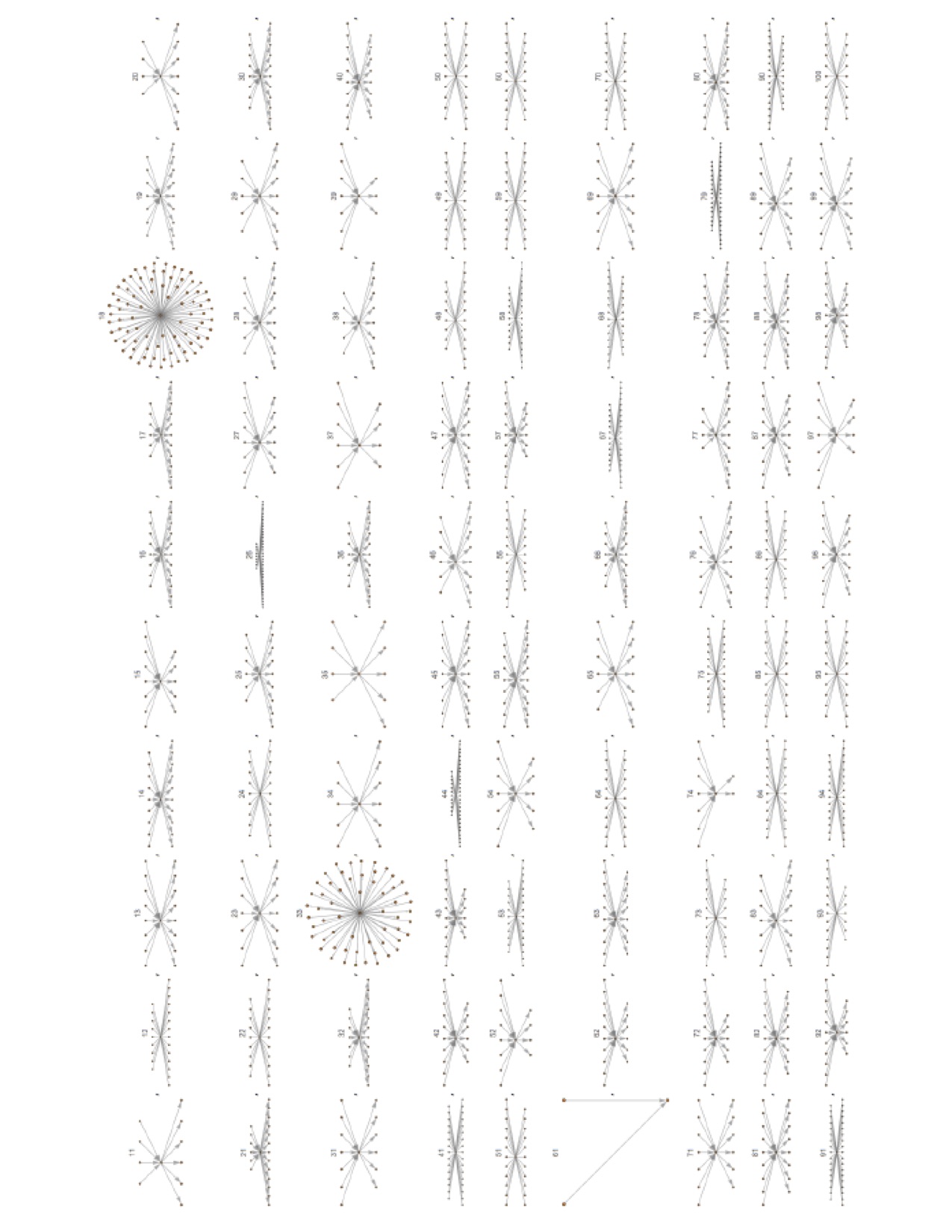}
\caption*{{\bf Figure S2}. Full set of motifs.}  
\end{figure*}

\begin{figure*}
\centering
\includegraphics[scale=.3]{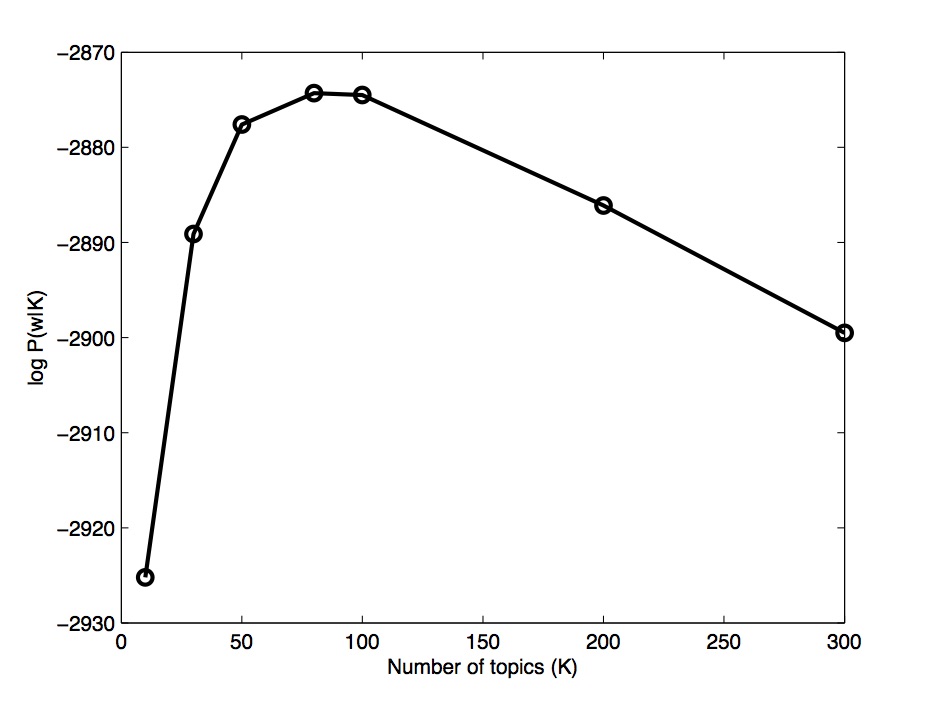}
\caption*{{\bf Figure S3.} Results of 5-fold cross-validation for model selection. The log-likelihood of held-out data is showed.}
\label{fig:lda_cv}
\end{figure*}

\begin{figure*}
\centering
\includegraphics[scale=.6]{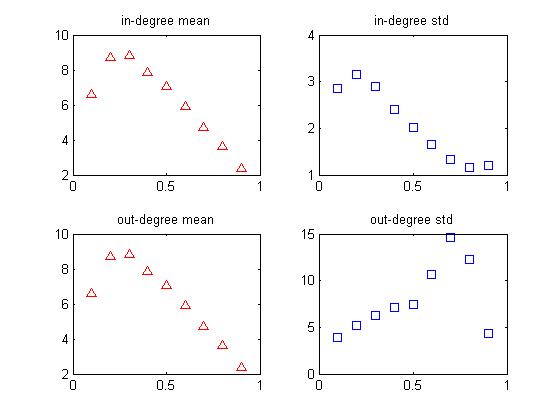}
\caption*{{\bf SI Figure S4.} The mean and standard deviation of the in and out degree of the inferred diffusion networks for the threshold parameter sampled between the values of $0$ and $0.8$. The density is non-monotonic and exhibits a maximum value at around 0.3. Threshold values toward zero yield disconnected networks and values toward 1 also generate sparse networks. For our analysis we select the threshold (0.3) that maximizes the density of connections. This is equivalent to maximizing variation through descent. }
\end{figure*}

\begin{figure*}
\centering
\includegraphics[scale=.4]{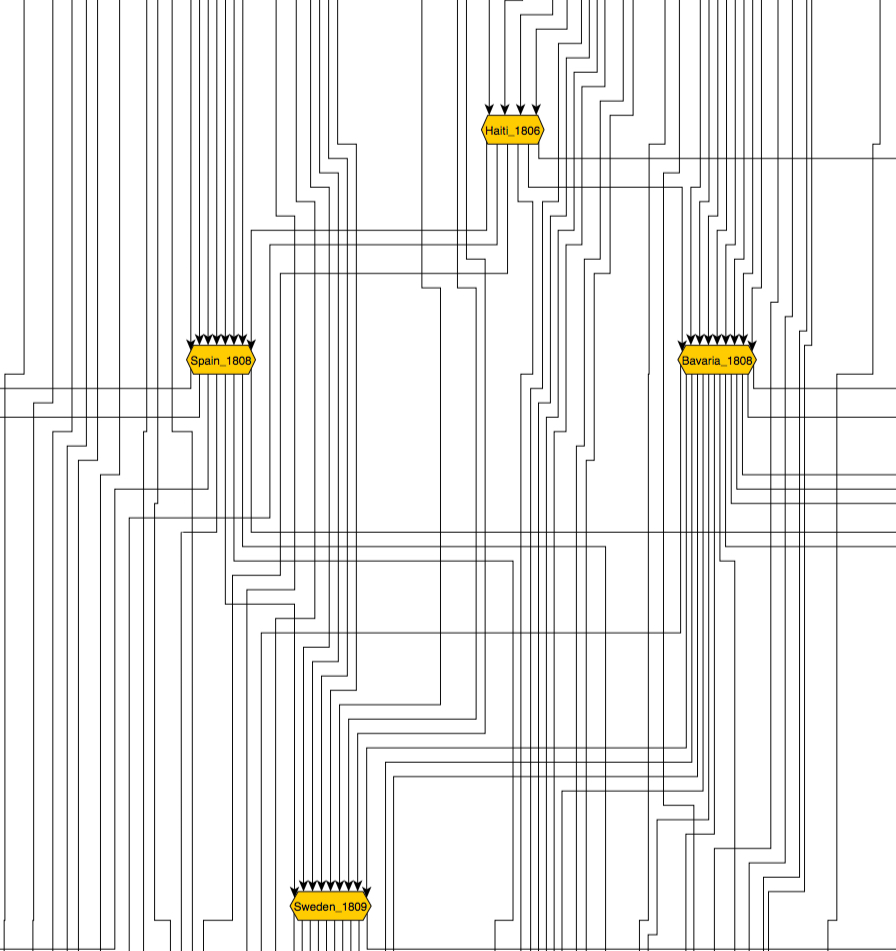}
\caption*{{\bf Figure S5.} Detail from the full diffusion network learned on the topic modeling of 591 constitutions. The arrows pointing in to a node indicate a topic expressed above some threshold earlier in the space of constitutions (a source) also being expressed above some threshold at this new point (the target).}
\end{figure*}
\begin{figure*}
\centering
\includegraphics[scale=.7]{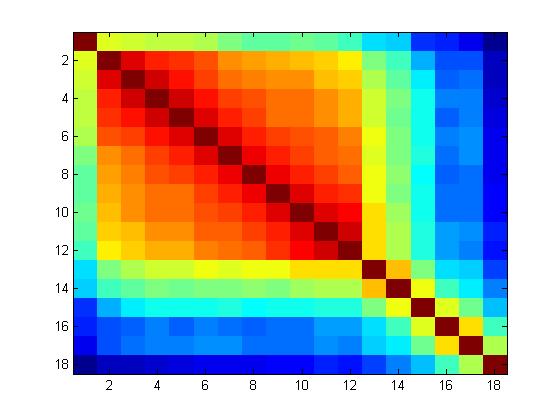}
\caption*{{\bf Figure S6.}  Indegree heatmap robustness. This illustrates how changes in the diffusion threshold have a negligible impact on the temporal ordering of  the indegree. In other words the number of constitutional parents of any given constitution is very stable with respect to our choice of threshold parameter. }
\end{figure*}

\begin{figure*}
\centering
\includegraphics[scale=.7]{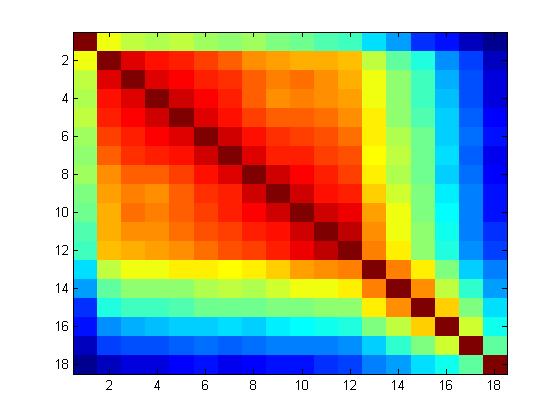}
\caption*{{\bf Figure S7.} Outdegree heatmap robustness. This illustrates how changes in the diffusion threshold have a negligible impact on the temporal ordering of  the outdegree. In other words the number of constitutional parents of any given constitution is very stable with respect to our choice of threshold parameter. }
\end{figure*}

\clearpage

\bibliographystyle{apacite}
\bibliography{template}
\end{document}